\documentclass[epj]{svjour}

\usepackage{amsmath}
\usepackage{epsfig}

\newcommand{\beq}{\begin{equation}}
\newcommand{\eeq}{\end{equation}}
\newcommand{\bea}{\begin{eqnarray}}
\newcommand{\eea}{\end{eqnarray}}

\begin{document}

\title{Demystifying generalized parton distributions}
\author{A.~Freund}
\institute{Institut f{\"u}r Theoretische Physik, Universit{\"a}t Regensburg, D-93040 Regensburg, Germany}

\date{\today}

\abstract{In this paper, I will explain in as simple and intuitive
  physical terms as possible what generalized parton distributions
  are, what new information about the structure of hadrons they convey
  and therefore what picture of the hadron will emerge. To develop
  this picture, I will use the example of deeply virtual Compton
  scattering (DVCS) and exclusive meson electroproduction processes.
  Based on this picture, I will then make some general predictions for
  these processes.
}

\maketitle

\section{Introduction}

Scientists have striven for centuries to unravel the dynamics and the
structures involved in the physical systems they have been
investigating, from large scale structures in our universe over
biological systems down to the smallest scales achievable in todays
high energy experiments. At these smallest scales the questions one is
trying to answer are ``What are the substructures of hadrons, what are
the dynamics of these substructures and what three dimensional picture
of hadrons is emerging ?''.

In the theory of strong or color interactions (QCD) parton
distribution functions (PDFs) encode the long distance or bound state
i.e.  nonperturbative information about hadrons. These PDFs are
precisely what we need in order to construct a dynamical as well as
geometrical picture of these objects.  Unfortunately, most high energy
experiments analyzing hadronic substructure study inclusive reactions
such as deep inelastic scattering (DIS) $e+p\to e+X$; in other words the
object they would like to study is destroyed in the reaction. Although
PDFs can be extracted from inclusive data, these functions are only
single particle distributions precisely because the target is
destroyed and, hence, they depend only on a longitudinal momentum
fraction, $x_{bj}$, and a transverse resolution scale, $\mu^2$. Since
inclusive PDFs do not contain information on the impact parameter of
the probe, vital information about the three dimensional distribution
of substructure is lost and, therefore, these PDFs can only give a one
dimensional picture of a hadron. It could be argued that so called
unintegrated PDFs (see for example \cite{ccfm}) contain more
information on hadrons, since the additional transverse scale can be
interpreted as a relative transverse position. However, this scale is
integrated over in physical observables and thus no direct information
can be deduced from it. In order to gain insight into a hadron's three
dimensional structure one has to measure particle correlation
functions which encode additional information on how the object as a
whole reacts to an outside probe in terms of physical observables.
Correlations in hadrons refer to the dynamical influence during the
reaction one or more particles or partons found in a particular state
inside the hadron have on one or more other partons found in a
different state inside the same hadron. A good example would be the
transition of a quark/gluon of a certain momentum into a configuration
inside the same hadron with a different momentum or the removal from
the hadron i.e.  transition into vacuum, of a $q\bar q$/gluon pair
with a certain momentum configuration.  Because of the closeness in
meaning between parton correlation and parton configuration, I will
use the two phrases interchangeably from now on.  Note that particle
correlation functions can only be measured if the hadron stays intact
during and after the reaction since the dynamical relationship between
the different partons would otherwise be destroyed.  This can only be
achieved if no large color forces, responsible for a break-up, occur
during the reaction.  This requirement forces such a reaction to be
mediated by color neutral objects such as color singlets or, at the
very least requires, that color is locally saturated.  The
experimental signature of such a process can be either a, so-called,
rapidity gap meaning that the produced particle/s which are well
localized in the detector, are clearly separated from the intact final
state hadron with no detector activity in between the two, or a small,
so-called, missing mass, which characterizes the difference between
the initial energy and the sum of the energies of all the
reconstructed particles in the detector.  There are many reactions of
this kind such as hard diffraction $e+p\to e+p+X$ or, in particular,
deeply virtual Compton scattering (DVCS) $e+p\to e+p+\gamma$
\cite{mrgdh,ji,ji2,rad1,jcaf,diehl97} which is the most exclusive
example of hard diffraction. Hard is meant here in the sense of the
presence of a large scale in the reaction such as a large momentum
transfer from probe to target.  In the perturbative QCD description of
fully exclusive hard reactions such as DVCS, we finally encounter the
objects we have been looking for: particle correlation functions. They
appear in the collinear factorization theorems of these reactions
\cite{jcaf,jfs} where collinear refers to the physics being dominated
by what is happening on the light cone neglecting internal transverse
momenta.  Factorization theorems state that, within QCD, one can
factorize the leading term in the cross section or scattering
amplitude of a particular hard reaction to all orders in perturbation
theory into a convolution of a finite or infra-red safe, hard
scattering function and an infrared sensitive, nonperturbative
function, a PDF. The remaining terms in the cross section or amplitude
are suppressed in the large scale of the reaction and can be
disregarded, at least in the limit of very large scales. The hard
scattering function is particular to each reaction but computable to
all orders in perturbation theory.  The PDFs which are universal
objects and can be used in other hard, exclusive reactions, cannot be
computed within perturbative QCD save for their momentum scale
dependence induced by the renormalization of the theory. They are
given, in a quantum field theoretic language, as a Fourier
transformation of a matrix element of non-local, renormalized,
operators. The key thing, in this context, are the in and out states
of these matrix elements. In inclusive reactions such as DIS, the in
and out states are the same since the scattering amplitude can be
directly related through the optical theorem to a reaction which has
the same in and out state. In hard, exclusive reactions the in and out
state differ, at least, in their momenta.  This is due to a finite
momentum transfer in the $t$-channel of the reaction onto the outgoing
hadron, most commonly a nucleon.  These PDFs depend on more variables,
namely those characterizing the momentum difference of the in and out
state, than the PDFs in inclusive reactions which only depend on one
momentum variable, apart from the momentum scale dependence and
therefore carry only one dimensional information on the hadron. The
behavior of these PDFs called generalized parton distributions (GPDs)
\cite{mrgdh,ji,ji2,rad1,barloew,poly} under a change of their
variables encodes the response of the entire hadron, i.e. its
substructure, to the outside probe.  Therefore, these GPDs are
particle correlation functions, more precisely light cone particle
correlation functions, and a complete mapping in all their variables
through experiments would give us for the first time a full three
dimensional picture of hadrons. Please note here that GPDs are by no
means the only particle correlation functions encountered in high
energy reactions. For example, so called, higher twist matrix elements
in DIS, which contain more than just two elementary operators, are
correlation functions since the momenta of the third, fourth etc.
operator in the matrix element depend on the momenta of the other
operators involved.  Furthermore, generalized distribution amplitudes
\cite{afex,polyrev,diehl2} encountered in exclusive $\gamma\gamma^*$ reactions
or transition GPDs in, for example, $e+p\to e+n+\pi^+$ are also
correlation functions (for a review see \cite{polyrev} and references
therein). Since the aim of this paper is not completeness but rather
an intuitive understanding of at least some of the physics involved,
we will only discuss afore mentioned GPDs and their physical
implications.

\begin{figure}
\centering
\mbox{\epsfig{file=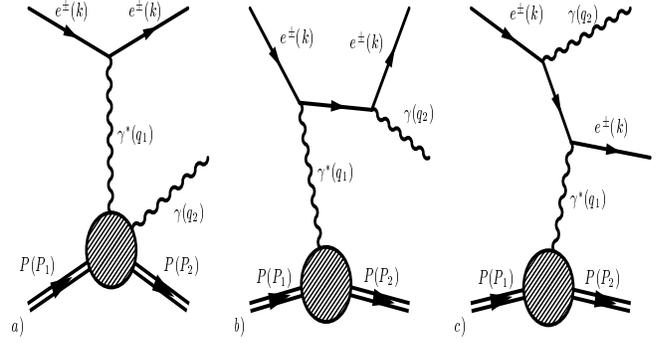,width=8.5cm,height=4.5cm}}
\vskip+0.2in
\caption{a) DVCS graph, b) Bethe-Heitler with photon from final state lepton
and  c) with photon from initial state lepton.}
\label{dvcsfig}
\end{figure}

In order to directly extract GPDs from experiment one has to access
scattering amplitudes. Unfortunately, the cross section for exclusive
processes is the amplitude times its complex conjugate, $|A|^2$,
compared to inclusive processes where the cross section is just given
by the imaginary part of the amplitude. Though we are accessing both
the real and the imaginary part of the amplitude in exclusive
processes, their phase structure i.e. each part individually, cannot
be cleanly separated unless there is a ``phase filter''.  A "phase
filter" would be a well understood process with which the exclusive
reaction interferes. Fortunately, there is such a process in the case
of DVCS, the QED Compton or Bethe-Heitler (BH) process (see Fig.\ 
\ref{dvcsfig}), first discussed in \cite{bordclose}. The interference
term between the two processes allows one to directly access both the
imaginary and the real part of the DVCS scattering amplitude which
contain, {\it simultaneously}, four distinct structures, namely ${\cal
  H}$, an unpolarized amplitude with no hadron spin-flip, ${\cal
  \tilde H}$, a polarized amplitude with no hadron spin-flip, ${\cal
  E}$, an unpolarized amplitude with hadron spin-flip and ${\cal
  \tilde E}$, a polarized amplitude with hadron spin-flip.  The
imaginary part is accessible through the measurement of the beam spin
asymmetry (longitudinal polarization in and opposite to the beam
direction) also called single spin asymmetry (SSA) and the real part
through the beam charge asymmetry (reversal of the lepton charge) or
simply called charge asymmetry (CA) \cite{diehl97,afmmlong,bemu4}.
This ``filtering'' has been aptly named ``nucleon holography'' by the
authors of \cite{bemunew}, since it employs the same principle of
interference as regular holography. Note that the nucleon spin-flip is
only made possible because of a finite momentum transfer $t$ onto the
final state nucleon as compared to DIS where $t=0$ and thus there is
no spin flip. This last statement means that ${\cal E}$ and ${\cal
  \tilde E}$ have no inclusive analog and hence contain unique
information on the nucleon only accessible in exclusive reactions.

Note, furthermore, that whereas on the amplitude level we have
``nucleon holography'', on the deep structure level of the GPDs we
will, as I will explain in a later section, have ``nucleon
tomography'' \cite{fps} (see also \cite{ral1}), since for each value
of $x_{bj}$ and each value of $t$ we are studying the dynamics of a
slice of a nucleon and so, when we put all of the slices together we
obtain a three dimensional image of a nucleon, as one obtains a three
dimensional image of a person when putting enough MRI pictures
together.

In Sec.\ \ref{picture}, I will define GPDs and then develop a picture
of what they mean in an intuitive way based on the example of DVCS and
exclusive meson production. In Sec.\ \ref{prediction}, I will make
general predictions about DVCS in particular and hard exclusive
reactions in general at facilities such as the planned EIC at BNL, the
proposed HERA III or a dedicated fixed target experiment. I will then
conclude in Sec.\ \ref{conc}.

\section{What is the physical picture GPDs convey?}
\label{picture}

\subsection{GPD Definition}

Whenever I will talk about GPDs in the following, I will refer to GPDs
in a nucleon, since I will mainly concern myself with hard
electroproduction reactions involving protons. However, the statements
below are much more general in nature and apply to any hadron target.
For brevity and ease of presentation, I will restrict myself to
nucleons.

GPDs, first implicitly introduced in \cite{mrgdh} and later
rediscovered in \cite{ji,rad1}, are generally defined through the
Fourier transform of matrix elements of renormalized, non-local
twist-two operators.  Twist-two operators are composite operators
containing only two elementary fields of the theory. These are
situated at different positions on a light ray making them non-local
and are sandwiched between {\it unequal} momentum nucleon states. The
essential feature of such light cone parton correlation functions,
where the difference in the in and out state is responsible for the
correlations, is the presence of a finite momentum transfer, $\Delta =
p-p'$, in the $t$-channel ($p, p'$ are the initial and final state
nucleon momenta). Hence, the partonic structure of the nucleon is
tested at {\it distinct} momentum fractions.

There are many representations of GPDs (see
\cite{mrgdh,ji,rad1,gb1,ffgs}).  In this paper I will use the
off-diagonal PDFs, ${\cal F}^i (X,\zeta)$, defined by Golec-Biernat and
Martin \cite{gb1} and used in the numerical solution of the
renormalization group or evolution equations in \cite{afmmgpd} (for
other treatments see \cite{gb1,ffgs,rad3,bemuevolve}).  This
representation will allow us a very intuitive insight into GPDs as I
will explain now.
 
The GPDs in this representation depend on the momentum fraction $X \in
[0,1]$ of the {\it incoming} proton's momentum, $p$, and the
skewedness variable $\zeta = \Delta^+/p^+$ (so that $\zeta = x_{bj}$ for DVCS and
meson production). This is analogous to the case of forward PDFs where
$x_{bj}$ is also defined with respect to the incoming proton's
momentum.

\begin{figure} 
\centering 
\mbox{\epsfig{file=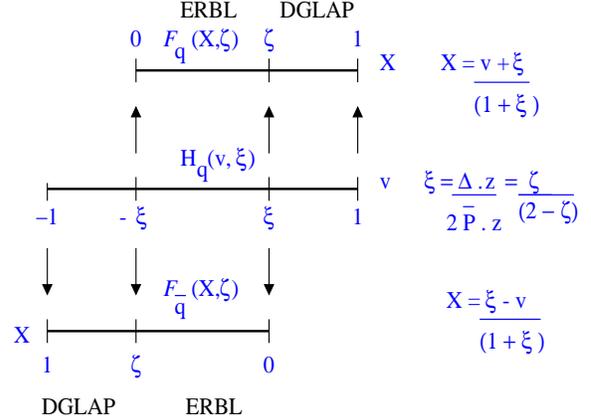,width=7.5cm,height=5.5cm}} 
\vskip+0.2in
\caption{The relationship between ${\cal F}^q (X,\zeta)$, ${\cal F}^{\bar q} (X,\zeta) $ and $H^q (x,\xi)$ with $x \in [-1,1]$ and $X \in [0,1]$.} 
\label{figfdd} 
\end{figure}

For the quark case, the relationship of the quark and anti-quark
distributions, ${\cal F}^q (X,\zeta), {\cal F}^{\bar q} (X,\zeta) $, to the
more widely used $H^q (x,\xi)$ \cite{ji} where the GPDs are defined with
respect to the average of $p$ and $p'$ ($x \in [-1,1]$ and $\xi = \zeta/(2-\zeta)
\in [0,1]$) is shown in Fig.~\ref{figfdd}.  More explicitly, for $x \in
[-\xi,1]$:
\begin{equation}
{\cal F}^{q,a} \left(X = \frac{x+\xi}{1 + \xi},\zeta\right) = \frac{H^{q,a} 
(x,\xi)}{1-\zeta/2} \, ,
\label{curlyq}
\end{equation}
\noindent and for $x \in [-1,\xi]$
\begin{equation}
{\cal F}^{{\bar q},a} \left(X = \frac{\xi -x}{1 + \xi},\zeta\right) = 
-\frac{H^{q,a} (x,\xi)}{1-\zeta/2} \, .
\label{curlyqbar}
\end{equation}

The two distinct transformations between $x$ and $X$ for the quark and
anti-quark cases are shown explicitly on the left hand side of
eqs.(\ref{curlyq},\ref{curlyqbar}). There are two distinct regions:
the DGLAP region, $X > \zeta$ ($|x| > \xi$), in which the GPDs behave like
regular parton distributions and obey a generalized form of the so
called DGLAP equations for PDFs, and the so called ERBL region, $X<\zeta$
($|x| < \xi$), where the GPDs behave like distributional
amplitudes/meson wavefunctions and obey a generalized form of the ERBL
equations for distributional amplitudes (see
\cite{gb1,ffgs,afmmgpd,rad3,bemuevolve}). In the ERBL region, due to
the fermion symmetry, ${\cal F}^q$ and ${\cal F}^{\bar q}$ are not
independent anymore. In fact ${\cal F}^{q} (X,\zeta) = -{\cal F}^{\bar q}
(\zeta-X,\zeta)$, which leads to an anti-symmetry of the unpolarized quark
singlet distributions (summed over flavor $a$), ${\cal F}^S = \sum_a
{\cal F}^{q,a} + {\cal F}^{{\bar q},a}$, which is C-even, about the
point $\zeta/2$ (the C-odd non-singlet and the C-even gluon, ${\cal F}^g$, which
is built from $x H^g(x,\xi)$, are symmetric about this point). For a
detailed review of the mathematical properties see, for example,
\cite{ji2}.

The operator definition of the ${\cal F}$'s is analogous to the one
for the $H$'s:
\begin{align}
&{\cal F}^q(X,\zeta) =\nonumber\\
&\int \frac{dz^-}{4\pi}~e^{-i(X-\zeta)p^+z^-}\langle 
p|{\bar \psi}\left(z^-\right){\cal P}\gamma^+\psi\left(0\right)|p'\rangle\nonumber\\
&{\cal F}^g(X,\zeta) =\nonumber\\
&\int \frac{dz^-}{2\pi Xp^+}~e^{-i(X-\zeta)p^+z^-}\langle 
p|G_{+\nu}\left(z^-\right){\cal 
P}G^{\nu}_+\left(0\right)|p'\rangle
\label{gpddef}
\end{align}
except that the Fourier conjugate momentum fraction, the light cone
positions and the momenta of the in and out states are different
compared to the symmetric approach. Note that one could have, more
conventionally, chosen $X$ to be the Fourier conjugate momentum to
$z^-$.  Since the crucial points $X=0$ and $X=\zeta$ are related via the
above symmetry arguments, it does not matter whether one chooses one
or the other. Nonetheless, the variable $X-\zeta$ will prove convenient
later on since it will be zero for $X=\zeta$ which is a special point and
signals that large, strictly speaking infinite, light-like separations
of the operators will play a very important role in the GPD. As we
will see below, this point in the GPD is of paramount importance in
hard exclusive reactions like DVCS and meson production. In the
symmetric representation \cite{ji}, the uniqueness of this point in
terms of separation of operators on the light ray is not as obvious
and thus I prefer a representation here where the uniqueness of this
point is directly apparent. This does not mean that one representation
is better than another but rather that sometimes one representation is
more convenient to use than another.

Below, I will refer quite often to valence and sea quark
distributions. In terms of eq.~(\ref{gpddef}) the valence or
C-odd non-singlet quark distribution of a flavor $a$ is defined as
\begin{equation}
{\cal F}^{a}_{val} =  {\cal F}^{q,a}-{\cal F}^{\bar q,a}
\label{defval}
\end{equation}
such that the first moment in $X$, summed over all flavors, yields the
number of quarks in the proton, and the definition of singlet
quark distribution for a given flavor $a$, a C-even GPD combination, is
\begin{equation}
{\cal F}^{S,a} =  {\cal F}^{q,a}+{\cal F}^{\bar q,a}
\end{equation}
which gives the sea quark distribution of flavor $a$
\begin{equation}
{\cal F}^{q,a}_{sea} = \frac{1}{2}\left[{\cal F}^{S,a}-{\cal F}^{a}_{val}\right]={\cal F}^{\bar q,a}_{sea}.
\label{defsea}
\end{equation}
Note that in the ERBL region, as pointed out above, the quark and
anti-quark distributions are not independent from one another anymore
and one can only speak of non-singlet and singlet distributions per
flavour $a$ without being able to separate out the sea.

\subsection{Why does DVCS help us understand GPDs better?}
\label{dvcsfirst}

The first question one has to answer is: Why is it that DVCS (see
Fig.\ \ref{dvcsfig}) is the cleanest process within which to measure
GPDs in a nucleon? The reason for this is quite simple. With a real
photon, one has an elementary, point-like particle in the final state
rather than a bound state like a meson or an even more complicated
state like several mesons/hadrons or jets adding other unknown,
nonperturbative functions.  Note that the contribution of the
non-point-like part of the real photon wave function which is similar
to a meson wavefunction, is power suppressed in DVCS \cite{rad1,jcaf}.
The factorization theorem for the DVCS scattering amplitude
\cite{rad1,jcaf} is merely a simple convolution of a hard scattering
function with only {\it one} GPD rather than with a GPD plus another
nonperturbative function as in meson production. To be more precise,
DVCS is only sensitive to a charge weighted C-even GPD combination
($\sum_ae^2_a{\cal F}^{S,a}= \sum_a e^2_a( {\cal F}^{q,a}+{\cal F}^{\bar
  q,a})$) in leading order (LO) of perturbation theory (the gluon GPD
enters only in next-to-leading order (NLO)), which is the flavour sum
over the singlet quark distribution for a given flavor $a$. Hence,
DVCS does not discriminate between different quark flavours as for
example exclusive $\pi^0$ production does due to its quark content
specific final state.

\begin{figure}
\centering
\mbox{\epsfig{file=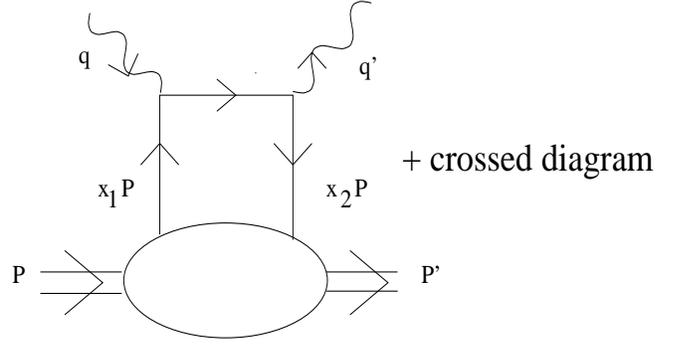,height=4.5cm,width=8.5cm}}
\vskip+0.2in
\caption{LO handbag diagram for DVCS. Here $x_1=X$ and $x_2=X-\zeta$.}
\label{hand}
\end{figure}

The DVCS amplitude is ${\cal T}\simeq Im{\cal T}\propto \sum_a e^2_a{\cal F}^{S,a}(\zeta,\zeta,Q^2)$ in
LO (see for example \cite{mrgdh,ji,rad1}).  This is true up to a
$\zeta=x_{bj}\simeq 0.2-0.3$ even when taking NLO effects into account
\cite{afmmlong,bemu4,bemu2,afmmamp}. Hence, DVCS is dominated, at
least in a very broad region of phase space, by the crossover point
between the DGLAP and ERBL region. At this particular point in phase
space, $X=\zeta$, the parton line carrying momentum fraction $x_2$ in
Fig.\ \ref{hand} is becoming ``soft'' and all the momentum is carried
by the incoming quark with fraction $x_1=X=\zeta$.  Also note that the
quark connecting the two photon vertices, which is usually hard i.e.
has large virtuality, is on or almost on mass shell and carries only a
large - momentum (see again Fig.\ \ref{hand}).  Factorization for DVCS
still holds in this situation \cite{jcaf}, with the hard interaction
now being the photon-quark vertex, however, the point $X=\zeta$ in the GPD
is indeed rather peculiar. One should recall that the GPD is defined
by a Fourier transform of a non-local matrix element on a light ray
and that the Fourier conjugate variables are the light-ray separation
$z^-$ between operators and a momentum fraction variable.  Here this
is either $X$ or $X-\zeta$ (see eq.\ (\ref{gpddef})). This means then that
for $X-\zeta\to 0$ , $z^-\to\infty$ and therefore the operators have an infinite
separation on the light ray or more physically speaking that there is
bad resolution of the probed object in the - direction on the
light cone.  This situation is analogous to inclusive DIS in the limit
of $x_{bj}\to0$.  Thus inclusive scattering at small $x_{bj}$ and DVCS
up to a large $x_{bj}$ in or, at least, near the valence region
(${\cal F}^{a}_{val}>{\cal F}^{a}_{sea}$) is dominated by the same
type of particle configurations with the only difference being that
the configurations in DVCS remain correlated since the proton stays
intact! What does the last statement mean from a physical point of
view?

\subsection{The physical picture of DVCS and its connection to GPDs}
\label{dvcspic}

The answer to the last question in Sec.\ \ref{dvcsfirst} is simply:
The particle configurations/correlations dominating the DVCS cross
section are much bigger, in their extension on the light ray, than the
probed object itself. Since the produced photon is a point-like object
these particle configurations which one would normally call
``end-point'' contributions, are not suppressed as in, for example, a
meson wave function describing an object of ``finite'' size
\cite{meson}! This suggests that even in the valence region, one is
not probing the actual bound quark structure both valence and sea but
rather QCD vacuum fluctuations as influenced by and interacting with
this bound state quark structure.  By QCD vacuum fluctuations, I refer to the
existence of two separate contributions, a nonperturbative and a
perturbative one. The perturbative QCD vacuum fluctuations will be
discussed in detail in Sec.~\ref{origin} when I discuss the origin of
the dominant parton configurations in DVCS, and the nonperturbative
QCD vacuum fluctuations can best be described as the spontaneous
fluctuation of color fields into $q\bar q$ pairs as well as the
formation of topological non-trivial color field structures like
instantons \cite{instan}.

Note a caveat here, though: The operators are not literally separated
by an infinite light-like distance, this would only be true in the
limit $Q^2\to\infty$, but rather by a distance which is inversely
proportional to, at most, $X-\zeta = \zeta\frac{\Lambda^2_{QCD}}{Q^2}$
(see \cite{afmmshort}) which acts as a lower bound and is motivated by
considering the fact that the intermediate quark in Fig.~\ref{hand} is
not exactly on mass shell.  To be definite compare this to DIS at
$x_{bj} = 0.2$ and an initial, nonperturbative, scale
$Q^2_0=1~\mbox{GeV}^2$. $X-\zeta=X-x_{bj}$ would then be bounded by
$0.2\cdot(0.2)^2/1=0.08$, which is not too small but still $2.5$ times
smaller, and at $Q^2=5~\mbox{GeV}^2$, $12.5$ times smaller, than the
respective momentum fractions encountered in DIS. The basic claim is:
{\it DVCS probes a larger, light-like distance than DIS for the same
  $x_{bj}$.}

\begin{figure}
\centering
\mbox{\epsfig{file=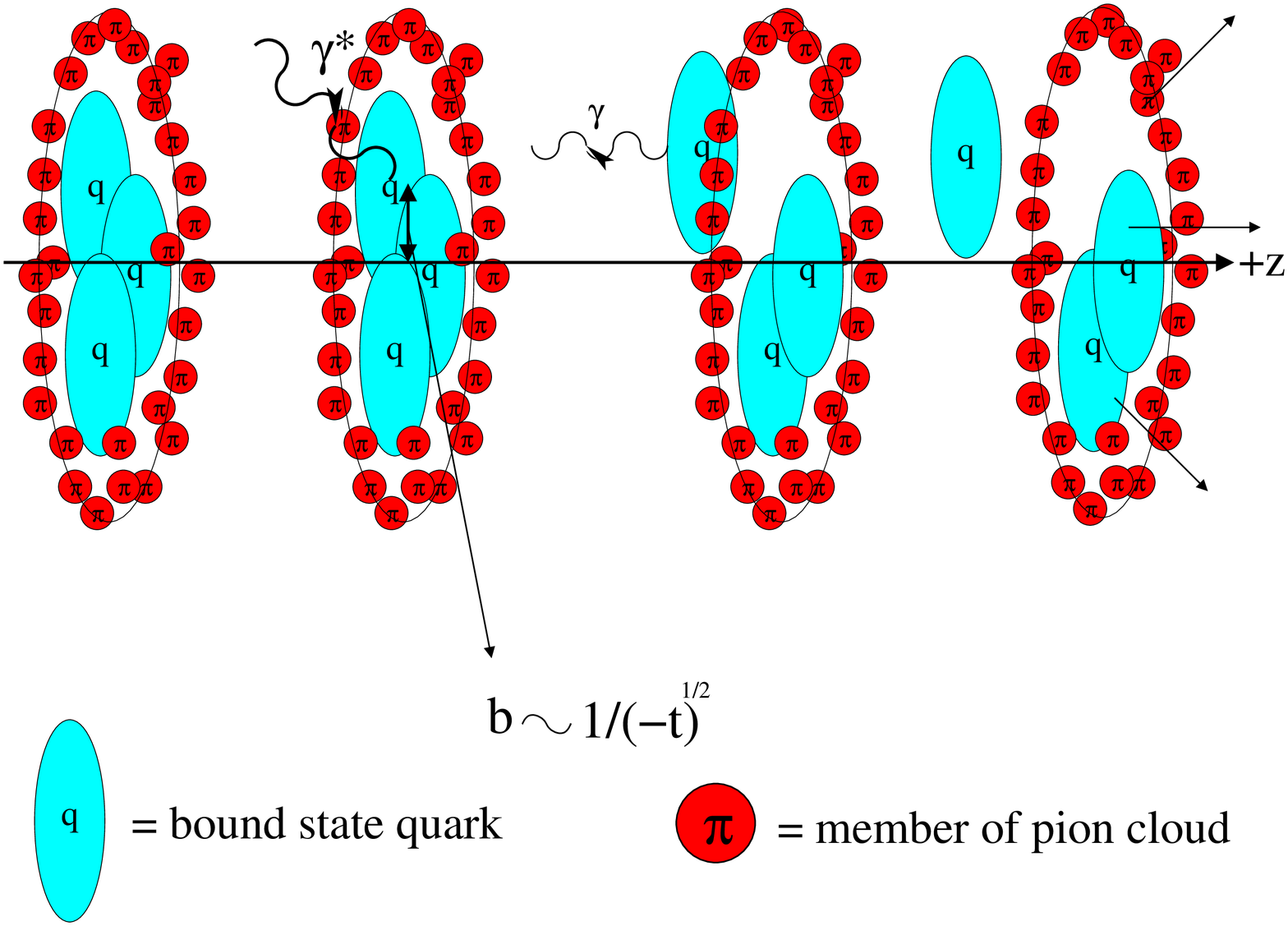,width=8.5cm,height=8cm}}
\caption{Stylized picture of how striking a valence quark and creating a real photon will lead to a proton breakup.}
\label{figproton}
\end{figure}

\begin{figure}
  \centering \mbox{\epsfig{file=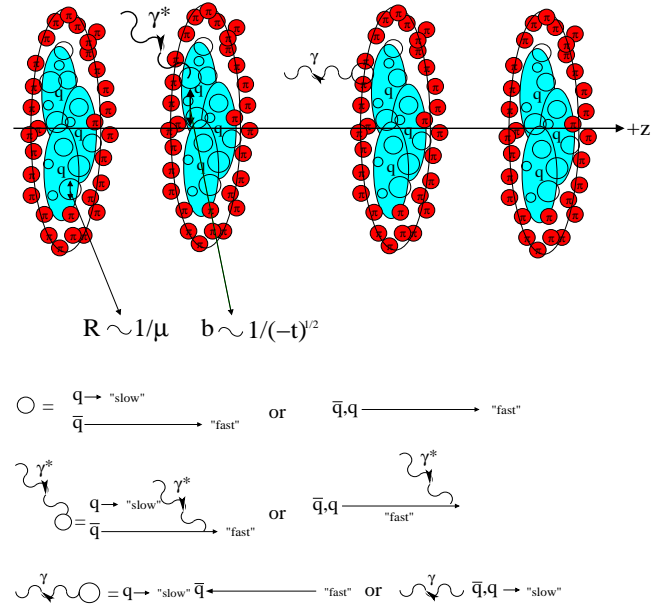,width=8.5cm,height=8cm}}
\caption{Stylized picture of how DVCS proceeds through sea configurations. The empty circle corresponds to possible partonic sea configurations through which DVCS can proceed, the empty circle struck by the virtual photon corresponds either of the previous configurations interacting with the virtual photon and the empty circle emitting a real photon corresponds to the possible creation mechanism through parton annihilation or radiation.}
\label{figproton1}
\end{figure}

There is a very intuitive picture of why the above interpretation is
indeed true and one is not really probing the actual nonperturbative
bound state structure structure of the proton within DVCS but rather
the quark and gluon configurations which are not relevant for the
bound state.  Consider the following situation (Fig.\ \ref{figproton})
at a low momentum scale $Q$: In the infinite momentum frame, the
proton is moving along the $+$ direction of the light cone i.e. in the
positive $3$ or + $z$-direction with each bound state quark carrying
on average a momentum fraction, $X\simeq O(0.1)\simeq x_{bj}$. If such a quark
were to be struck by a virtual photon which has large + $-x_{bj}P_+$
and $-Q^2/2x_{bj}P_+$ components with $P_+ \simeq O(Q)$, it would then only
have a large - component but a quasi zero + component since $X\simeq
x_{bj}$.  This means that the struck quark would have a large momentum
in the $-z$-direction, opposite to that of the other two quarks, then
radiate a real photon which moves in the $-z$-direction.  After
radiating the photon, the quark will then become ``soft'' i.e.  has no
large momentum components. The transition matrix element i.e.  the
overlap integral, between an initial state with $n$ \cite{foot3} bound
collinear or ``fast'' quarks to a final bound state with $n-1$
collinear quarks and one soft or ``slow'' quark is suppressed.  This
is due to the probability of two collinear and one ``soft'' quark
forming a proton in the final state being linearly suppressed with the
relative light-like separation or in momentum space with the momentum
fraction, $X-\zeta$, of the ``slow'' quark (see \cite{dfjk} eq. (53)).
DVCS, however, is observed at large $x_{bj}$ \cite{hermclas} and low
$Q$, therefore, the only alternative picture (Fig.\ \ref{figproton1})
is the one where the virtual photon is not scattering on bound state
quark but rather on a $q/\bar q$ from sea configurations/QCD vacuum
fluctuations which are {\it not} relevant for the actual bound state.
In these configurations the $q/ \bar q$ has a large + momentum
fraction which is are at large $x_{bj}$ thus making DVCS rare,
matching the one from the virtual photon.  In other words, the struck
$q/ \bar q$ starts to move in the -z direction and then annihilates
with a ``soft'' ($X-\zeta\simeq0$) $\bar q/ q$ from the sea into a real photon
or radiates a real photon and becomes ``soft''.  This real photon has
large - momentum i.e.  moves along the -z direction as it should.
None of the bound state quarks are directly involved in the reaction
and therefore, it is not very difficult for the proton to stay intact.
This statement can be equivalently recast in saying that the physics
of the bound state itself is not disturbed by the reaction. This
implies that the bound state quarks themselves will mainly be found in
symmetric configurations as in DIS. Also note that for the asymmetric,
``fast'' $\to$ ``slow'', configurations above, there will be no large
color forces since color is conserved locally through either event.
The above has consequences for the quark GPD with its valence and sea
part. For $X\sim\zeta$, where the configurations are asymmetric, the
nonperturbative valence distribution will be suppressed compared to
the inclusive case as well as the unknown part of the nonperturbative
sea necessary for the bound state.

The inclusion of gluons (their contribution is suppressed by $\alpha_s$)
does not change the above developed picture and interpretation. To
produce the required asymmetric gluon configuration, the collinear
gluon, with + momentum $X\simeq O(\zeta)$, has to split into a $q\bar q$ pair: a
hard $q$/$\bar q$ with large + and transverse momentum, interacts with
the $\gamma^*$, after which it remains hard but now with large $-$ rather
than + momentum ,and then annihilates with the other hard $\bar q$/$q$
which has only large negative transverse momentum, into a real photon
with only large $-$ momentum. The soft gluon $X-\zeta\simeq O(0)$ for the color
matching of the collinear gluon can be absorbed/radiated from either
$q$ or $\bar q$. This will leave the proton intact since, once more,
the bound state quarks are not directly involved and color is locally
conserved.
The reader might wonder why gluons with $X>>\zeta$ seem not to contribute
to the imaginary part of the amplitude, even though formally they do?
The answer to this question is an empirical one. Formally the
imaginary part of the gluon amplitude is given by
\begin{align}
&\mbox{Im}~{\cal T}^{g,V/A}_{DVCS} (\zeta,Q^2) = \frac{1}{N_f} \left(\frac{2 - \zeta}{\zeta}\right )^2  \Bigg[  \nonumber\\ 
&\int^1_{\zeta}dX\Big[\mbox{Im}T^{g,V/A} \left(z\right)\left({\cal
F}^{g,V/A} (X,\zeta)-{\cal F}^{g,V/A} (\zeta,\zeta) \right) \Big] \nonumber\\ 
&+ {\cal F}^{g,V/A}(\zeta,\zeta) ~\mbox{Im} \int^1_0 dX~T^{g,V/A} \left(z \right) \Bigg]
\label{imgamp}
\end{align}
with a an identical structure for the quark part. Note that the second
term in eq.~(\ref{imgamp}) is proportional to the gluon GPD at the
point $\zeta$ and this second term is usually the dominant contribution up
to a $\zeta \simeq 0.1$. Furthermore, in the integral of eq.(\ref{imgamp}), the
region $X\simeq O(\zeta)$ does, contrary to expectations, contribute a fairly
large part to the value of the integral. In consequence one can indeed
say that for small to medium $\zeta$ the simplified picture from above is
indeed the correct one.

For small $x_{bj}$, the picture does only change in so far as that
there are now no bound state quarks anymore which are ``visible'' to
the probe and DVCS definitely has to proceed via the above advocated
asymmetric parton configurations which, at a nonperturbative scale,
should be mainly found in the sea.

There are two things to note here, first, the above mentioned sea
configurations will be very rare at low $Q$ and any $x_{bj}$ making
DVCS a rare event compared to DIS and secondly, that these sea
configurations cannot directly be identified in inclusive DIS, since
this would require too large a light-light separation as compared to
the one allowed in DIS.

Thus, one can conclude that {\it first, there exist asymmetric parton
  configurations/correlations in the proton, not directly associated
  with the bound state structure of the proton.  These parton
  correlations themselves are encoded in GPDs in the region around
  $X\simeq\zeta$ and can only be probed in hard exclusive reactions like DVCS}.

What happens at larger $Q^2$? Is the above picture still valid?

\subsection{The origin of the asymmetric parton correlations}
\label{origin} 

The final questions of the previous subsection are easily answered
when considering the perturbative evolution of GPDs as $Q^2$
increases: Perturbative evolution i.e.  the change of the GPD under a
change in the renormalization or momentum scale, strongly enhances the
$X\simeq\zeta=x_{bj}$ region in the quark singlet GPD. Within the singlet, the
sea is much more enhanced than the valence part, as compared to the
evolution effect in forward PDFs at the same $x_{bj}$ (see for example
\cite{ffgs,afmmgpd} for a detailed analysis of this phenomenon). This
enhancement effect is driven by the gluon GPD which itself is not as
strongly enhanced as the quark GPD, and the structure of the
perturbative evolution kernels \cite{bfd} favoring splitting into
asymmetric configurations. This is similar to the inclusive case where
the gluon PDF drives the rise of the quark sea, however not as strong
as in the GPD case at $X\simeq\zeta$. Note that the gluons responsible for the
enhancement at higher $Q^2$ originate themselves from quarks at higher
values of $X$ and lower values of $Q^2$ i.e. are collinearly radiated
from the nonperturbative i.e. low scale valence quarks in the proton
which are found {\it not} in asymmetric configurations but rather in
symmetric ones as encountered in DIS.  Thus evolution creates more and
more asymmetric correlations inside the proton as it transitions from
$\langle p|$ to $|p'\rangle$ and hence the valence quarks at low $Q^2$ become more
and more ``dressed'' at higher $Q^2$ (see Fig.\ \ref{figproton1}) or,
equivalently, their correlated, perturbative substructure, the sea or
perturbative QCD vacuum fluctuations, is more and more revealed as the
scale is increased (see Sec.~\ref{taqdep}) making DVCS more and more
likely without having to change the actual reaction mechanism.  One
can say then that at low $Q^2$ and large $X\neq\zeta$ (disregarding the $t$
dependence for the time being) the quark GPD and the inclusive quark
PDF should be the same since they are both dominated by the same type
of symmetric configurations at low $Q^2$.  However, since the
evolution is different for the GPD and PDF the two will be different
at higher $Q^2$.  In summary, {\it at large $Q^2$ and any $x_{bj}$ the
  asymmetric parton correlations responsible for facilitating DVCS
  are almost exclusively perturbative in nature}.

The question of the origin of the asymmetric parton correlations at
low $Q^2$ where perturbative evolution is either not valid anymore or
its use is questionable, is more difficult to answer. They should be a
nonperturbative feature of QCD vacuum fluctuations rather than the
valence structure which is found in much more symmetric configurations
as discussed above. Also, one might expect that nonperturbative
asymmetric configurations would be suppressed since they would look
like end-point configurations in a meson. At large $x_{bj}$ where the
valence quarks of the proton dominate, the expectation would be that
there are no or very few such asymmetric correlations (see Fig.
\ref{figproton}). However, at very small $x_{bj}$, when one enters the
high gluon density or non-linear regime, one might still be able to
answer the question from a perturbative point of view. This is true as
long as the natural scale of the problem is the so-called saturation
scale $Q_s=\left(\frac{x_0}{x_{bj}}\right)^{\lambda}\cdot Q_0$ with $\lambda\sim0.15-0.2$
and $x_0,Q_0$ some reference/normalization scales where the small $x$
evolution starts. This means that $Q_s$ will be large at small $x_{bj}$.
Saturation refers here to the effect that in the regime of large color
fields the overlap of gluon wavefunctions lead to destructive
interference effects which are characterized by essential
non-linearities in the relevant small $x_{bj}$ evolution equations for
the color correlators, for example dipoles (see for example
\cite{weigert} and references therein). These non-linearities slow down
the rapid increase of the number of gluons in the nucleon as $x_{bj}$
decreases.  This does not mean that the photon has virtuality $Q_s^2$
but rather that the internal scale of the gluon couplings in the
system is $\alpha_s(O(Q^2_s))$, which is small at sufficiently small $x_{bj}$
rather than $\alpha_s(Q^2)$ which at $Q^2 \leq 1~\mbox{GeV}^2$ is large.
This statement deserves a further explanation since it is counter
intuitive. It is most easily understood in the color dipole model (see
for example \cite{fgms1} and references therein) where the DVCS
amplitude is given as a convolution of a virtual photon wavefunction
with a dipole cross section and a real photon wavefunction. One can
easily show \cite{weigert} that the $Q^2$ dependence resides solely in
the wavefunction and that the dipole cross section depends only on
$x_{bj},x_0,Q_0$ i.e $Q_s$. The small $x$ evolution determines how
$\sigma_{dipole}$ changes as $x_{bj}$ decreases, {\it independent} of
$Q^2$. The scale $Q_s$ is determined by $\lambda$ which in turn is given by
the relative change in $ln(1/x_{bj})$ of the slope of the dipole
distribution in dipole size $r$ at the point where the distribution is
about $1/2$ (see \cite{weigert} and references therein). To be more
precise, in the evolution equation for the color correlator, which is
essentially $\sigma_{dipole}$, $\alpha_s$ appears underneath the convolution
integral of evolution kernel with a combination of linear and
non-linear color correlators. On inspection \cite{weirum,afhweg} it
turns out that the main contribution to the convolution integral stems
from dipole sizes of $O(1/Q_s)$ in the case of running coupling.
Contributions of larger dipoles (infra-red contributions) are
suppressed by the non-linearity (this is also true for fixed coupling)
and contributions from very small dipoles ($r \to 0, k_{\perp}\to \infty$,
ultraviolet contributions) are sufficiently suppressed due to the
smallness of $\alpha_s$. This is not the case, by the way for the fixed
coupling case. The key observation is therefore that the smallness of
$\alpha_s$, if $Q_s$ is large, allows a perturbative treatment of the
gluonic degrees of freedom and thus their evolution in $x_{bj}$,
despite that fact that the color fields are very large.  Note,
however, that this does not imply that the DVCS amplitude or the total
DIS cross section for that matter, is entirely perturbative at small
$x_{bj}$. But rather that the change in $\sigma_{dipole}$ with $x_{bj}$ is,
whereas the nonperturbative information at small $Q^2$ resides in the
photon wavefunction and in the initial condition for $\sigma_{dipole}$ at
$x_0,Q_0$.

In the above regime regime, one can therefore say that asymmetric
configurations again originate from perturbatively treatable gluon
configurations, as at large $Q^2$, though these configurations come
from completely different regions of phase space and thus correspond
to a different aspect of QCD vacuum fluctuations as compared to the
ones at large $Q^2$.  Let me add a note of caution here as far as the
identification of high density gluons with a gluon GPD is concerned.
The non-linear, small $x_{bj}$ evolution does not rely on a twist
expansion but rather includes {\it all} twists. In fact higher twist
contributions provide the essential non-linearities in the evolution
equations.

In summarizing one can say that {\it the main source of the asymmetric
  parton configurations are gluons originating themselves either from
  symmetric valence configurations at a lower scale or are part of
  nonperturbative QCD configurations at small $x_{bj}$}.

\subsection{Meson production and GPDs}

If one were to consider other reactions like meson production, the
situation, previously discussed, obviously changes since one does not
want to produce an elementary particle which is predominantly
point-like and therefore easily allows particle configuration of
``infinite'' extent in its creation but rather a bound state with a
``finite'' size.  As I will explain below, only some details are
adjusted, the overall picture, however, remains unaltered.

As in DVCS in LO of perturbation theory, the imaginary part of the
scattering amplitude in meson production is proportional to ${\cal
  F}(\zeta,\zeta,Q^2)$. Depending on the produced meson i.e. its
quantum numbers, a particular combination or particular types of GPDs
are probed in contrast to DVCS where only the quark singlet is
directly probed. Therefore, the mesons act as a ``GPD
filter''. For example $\pi^0$ production, being a pseudo scalar, singles
out the polarized quark GPD in an unpolarized reaction \cite{jfs}!

Consider the following picture of meson production, again in the
infinite momentum frame: The proton moves along the + direction of the
light cone and is struck by a highly virtual, longitudinally
polarized, so as to maintain factorization, photon again having large
+ and - light cone momenta. In order to produce a meson there has to
be the exchange of at least one gluon or equivalently the splitting of
a gluon into a $q\bar q$ pair. These can be, in keeping with the
factorization theorem for meson production \cite{jfs}, either hard or
collinear to the proton i.e. the $+$ direction, or collinear to the
produced meson i.e.  the $-$ direction. We are now particularly
interested in the situation when the struck collinear quark in the
proton (valence or not) carries the initial momentum fraction $X\simeq
x_{bj}$ with another accompanying quark/anti-quark being ``soft'' i.e.
$X-\zeta\simeq0$ as in DVCS.  This situation can only be achieved through the
exchange of at least one hard gluon. One can also probe the gluon GPD
directly, as in, for example, $J/\psi$ production \cite{jfs} at small
$x_{bj}$ where the gluon dominates. This corresponds to the case when
a collinear gluon carrying momentum fraction $X\simeq x_{bj}$, splits into
a $q\bar q$ pair, one of which interacts with the virtual photon and
the other one with a second, ``soft'', gluon. They then go on to form
the meson in the final state.  In both instances, one directly probes
the point $X=\zeta=x_{bj}$ in the GPD associated with a large light-like
separation of operators as in DVCS.

Let me discuss the quark case first and then speak about the gluon
case. When the collinear quark, which will eventually interact with
the virtual photon, radiates a hard gluon, the quark itself becomes
hard. We need the situation where the + component of the hard gluon is
small i.e. it is on or almost on mass shell, in exact analogy to the
quark connecting the two photon vertices in DVCS in Fig.\ \ref{hand}
for the situation $X\simeq\zeta$ as explained in Sec.~\ref{dvcspic}.  The quark
remains hard and at the photon-quark vertex, the struck quark starts
to move along the $-$ direction, since the + components of the virtual
photon and quark cancel. The gluon now splits either into a $q\bar q$
pair with the anti-quark carrying large $-$ momentum or it hits a
``soft'' anti-quark in the proton transferring its large $-$ momentum.
In both instances the soft quark will be associated with the proton.
In order to keep the proton intact, the struck collinear quark could
not have been a valence quark since there would be no other collinear
i.e.  ``fast'' quark to replace it, only a ``soft'' i.e.  slow one.
Thus it must have come from a non-valence like configuration leaving
only the sea. In this way, the situation is analogous to the DVCS
case.  Hence, the interpretation of the exact particle configurations
probed in the GPD in meson production compared to DVCS for $X\simeq x_{bj}$
does not change for the case of quark scattering. What happens when we
have a collinear gluon as mentioned above?

The situation is quite similar to the quark case. To produce an
asymmetric configuration, the collinear gluon has to split into a hard
$q\bar q$ pair where either the hard $q$ or $\bar q$ has to go on or
near mass shell only carrying large - momentum, implying that the
initial, collinear gluon has a + momentum fraction $X\simeq\zeta$, and then
radiating a ``soft'' gluon required to match the color of the initial
gluon. Again we have the same situation as in the quark case and
therefore the same interpretation except that we have now the gluon
GPD rather than the quark GPD at $X=\zeta$ as already stated above.  The
origin of these asymmetric gluon configurations is the same as the one
for the quark case as explained in Sec.~\ref{origin}.

The fact that the interpretation about dominant particle
configurations encoded in the GPD does not change in going from DVCS
to meson production means that GPDs are indeed universal objects as
already proven to all orders in the factorization theorems
\cite{jcaf,jfs}.  However, it is nice to see how this universality
emerges from the simple physical picture above. Again, I would like to
stress that this picture of dominant particle configurations in meson
production is only valid if the imaginary part of the scattering
amplitude is larger than the real part. In fact, for the real part
where the regions $X>>x_{bj}$ and $X<<x_{bj}$ are very important,
valence quarks and symmetric gluon configurations do play an important
role. This is due to the fact that the region of phase space, where
the exchanged gluon or $q$ and $\bar q$ is hard, becomes large. It is
also clear that, as the mass of the produced vector meson or $Q^2$
increases, it starts to act in a similar fashion to a point particle
i.e directly emerges from the hard scattering space-time point.

The above also shows that the questions one asks of the proton in DIS
and hard, exclusive reactions are different. In DIS, on the one hand,
one asks the question if there are partons with large or small
momentum fractions in the proton, in hard, exclusive reactions, on the
other hand, one asks the much more specific question of how the
partons in the proton must conspire to make the reaction happen and
therefore one obtains a much more specific answer. One can then
conclude that {\it there exist asymmetric parton
  configurations/correlations in the proton, the exact nature of which
  can only be probed in hard exclusive reactions like DVCS or meson
  production.  These parton correlations are encoded in a GPD in the
  region around $X\simeq\zeta$}.

\subsection{The $t$ and $Q^2$ dependence of GPDs}
\label{taqdep}

Up until now, I have neither talked about the role of the
$t$ dependence nor of the precise meaning of $Q^2$ or more precisely
the renormalization scale $\mu^2$. In \cite{diehl1} a beautiful
exposition of the physical meaning of these two variables for GPDs has
been given (see also \cite{burkhardt}) which I will only briefly
reiterate: The scale $\mu^2$ defines from what scale, or, in space-time,
from what resolution in the transverse plane, onwards one can speak of
several or just one parton.  In other words, the better the resolution
$1/Q\sim1/\mu$ of the probe, the more partons or substructure of one parton
one can observe (see Fig.\ \ref{figproton1}). As $\mu$ defines the
resolution of the parton in the transverse plane, the $t$ dependence
gives the relative transverse position of the probed parton
correlation with respect to the proton (see Fig.\ \ref{figproton}). If
$\mu^2\simeq-t=$ several $\mbox{GeV}^2$, the exact meaning between resolution
and position becomes lost, including the above simple picture of DVCS
and meson production, since the hierarchy of scales necessary for a
factorized approach to these processes is lost and hence also its simple
physical picture.

In contrast, in the case of $\mu^2=Q^2>>-t$, one has a very interesting
picture emerging (see Fig.\ \ref{figproton1}): since $t$ is up to
corrections of $O(M_N^2\zeta^2)$, which are very small, equal to
$-(p_{\perp}-p'_{\perp})^2$, the relative transverse momentum difference
between initial and final state, small $t$ corresponds to a large
distance in the transverse plane from the proton ``center'' and large
$t$ to a small distance. Here ``center'' is meant with respect to the
relative transverse positional difference between initial and final
state. The question which arises now is where, relative to this
``center'', the asymmetric parton correlations take place ?

As far as the perturbatively generated correlations having a resolved
size of $O(1/Q)$ in the transverse plane are concerned, they will take
place closer to the ``center'', since they are associated with the
valence quarks through evolution and those have to be situated well
within the proton radius, $r_p\sim1$ Fermi. The nonperturbative
correlations have to be more clearly separated from the ``center'' of
the proton. The reason for this lies in the very fact that they cannot
be associated with the bound state structure as shown above and
therefore they will be have to be situated in the ``pion cloud'', for
lack of a better word, at the ``edge'' of the proton.

The emerging three dimensional picture of the asymmetric parton
configurations as well as their symmetric ``parents'' can be stated as
follows: The asymmetric parton configurations necessary to facilitate
hard, exclusive reactions are basically located ``inside'' of the
proton as it transitions from $\langle p|$ to $|p'\rangle$ during the reaction,
with the nonperturbative configurations towards the edge and the
perturbative configurations more towards the ``center'' but very
spread out on the light cone.  For example, at the average $t$ in DVCS
on the proton of HERMES of about $-0.2~\mbox{GeV}^2$, these
configurations are located only about $0.4$ Fermi away from the
``center'', clearly ``inside'' the proton charge radius $r_p\sim1$ Fermi
(only for a $t < -0.04~\mbox{GeV}^2$ would they be located ``outside''
of the proton charge radius $r_p$). Since we restrict our
considerations to the region of $-t\leq 1~\mbox{GeV}^2$, the relative
distance to the ``center'' is never closer than about $0.2$ Fermi.

One can now also understand why the cross section of hard, exclusive
processes drops when $t$ is increased and how this depends on $x_{bj}$
and $Q^2$. To do this, consider the following (see Fig.\ 
\ref{figproton1}): At low $Q^2$ and fixed $x_{bj}$, the main source of
the asymmetric configurations will not yet be perturbative collinear
parton splitting as at large $Q^2$ but rather some nonperturbative
property of QCD vacuum fluctuations. This means that, at low $Q^2$, as one
approaches the ``center'' of the proton i.e. as $t$ increases, the
number of asymmetric configurations suitable to facilitate a hard,
exclusive event should drop since the nonperturbative configurations
sit at the edge rather than in the ``center'', while at the same time
the perturbative configurations as part of the substructure of the
valence quarks, are not as well resolved yet as at higher $Q^2$ and
hence less than at large $Q^2$. In consequence, the average number of
asymmetric configurations available to the reaction is less at larger
$t$ than at smaller $t$, and as a consequence, the cross section drops
faster with the increase in $t$ at low $Q^2$ than at large $Q^2$.
Furthermore, as $x_{bj}$ decreases i.e. the energy increases, the
number of gluons from which asymmetric correlations can originate will
also increase, since more and more gauge fields will become ``frozen''
in the light cone time $z^+$ (see \cite{weigert} and references therein
for details) and can thus serve as a source. This means that the cross
section will drop faster with increasing $t$ at larger $x_{bj}$ than
at smaller $x_{bj}$. Since the evolution in $x_{bj}$ is less dramatic
than in $Q^2$ (see again \cite{weigert} and references therein), the
effect on the $t$ dependence will be less.

These observations are borne out both by the observations made
in \cite{fmsnew} where a $Q^2$ dependent but basically $x_{bj}$
independent slope gives very good agreement, within the experimental
errors, between the DVCS data and NLO QCD calculations and by
experimental measurements (see for example \cite{brho} and references
therein).

\section{Going beyond the nucleon: Qualitative predictions from the above 
picture}
\label{prediction}

The above considerations are not limited to a nucleon target but are
also valid for example for a nuclear target. There are some
interesting consequences emerging from the above considerations: The
fact that the same large light-like distances are involved in
conventional PDFs for $x_{bj}\to0$ and in GPDs for $X\simeq\zeta$ together with
the observed enhancement of this region through perturbative
evolution, suggests that for the same $x_{bj}$ of the process, {\it
  GPDs probe the configuration content of the proton and its effect on
  the QCD vacuum at relatively smaller momentum fractions than PDFs}.
This is borne out by the analysis carried out in \cite{fmsnew} which
shows that a good GPD input capable of describing all available DVCS
data \cite{hermclas,h1,zeus2} in a NLO QCD analysis is obtained by
using conventional forward PDFs at a momentum fraction $X$ shifted to
a smaller value by an amount of $O(\zeta)$.  This in turn implies
\begin{itemize}
\item Earlier onset of saturation effects in DVCS observables
  dominated by the imaginary part of the scattering amplitude compared
  to inclusive observables. This is particularly true for nuclear
  targets since saturation is a strongly $x_{bj}$ dependent phenomenon
  \cite{sgb}! A concrete prediction would be the presence of geometric
  scaling in the $\gamma^*p$ DVCS cross section in either $ep$ or $eA$
  scattering up to an $x_{bj}$ where it normally would break down in
  $F_2^{p,A}$ \cite{munier}.
\item Nuclear shadowing corrections for DVCS should set in at larger
  values of $x_{bj}$ as compared to the inclusive case. Moreover, at
  comparable values of $x_{bj}$, the nuclear shadowing corrections
  should be stronger in DVCS compared to DIS. Since nuclear shadowing
  is only a weak function of $x_{bj}$ except for the transition region
  between $0.01<x_{bj}<0.1$ (see for example \cite{fgms}), the
  enhancement effect would probably be mainly visible in this
  region \cite{afms}.
\item Since varying $t$ changes the relative transverse position at
  which the target is probed, it will allow one to scan through the
  ``grey'', where non-linear perturbative QCD is still applicable, and
  the ``black'' or total absorption region of the target. In these two
  regions, the target behavior will be qualitatively different and
  this difference should be reflected in different geometric scaling
  curves for different values of $t$ \cite{munier}. I do not claim
  here that DVCS in the black disc limit is very different from DIS in
  this limit, quite on the contrary \cite{fgms1}. However, the $t$
  dependence allows one to discern between two regions of different
  target behavior.
\end{itemize}

These predictions could be verified at the future EIC with its high
luminosity both for $ep$ and $eA$ scattering, as well as at HERA III
with nuclei in the HERA ring or a dedicated, high luminosity, fixed
target experiment.

Furthermore, the fact that one cannot really probe the bound state
quark distributions at $X=\zeta$ and leave the proton intact, leads one to
conclude that as $X\to\zeta$ the nonperturbative unpolarized valence quark
GPD should become small relative to the inclusive valence PDF at the
same $x_{bj}$ or tend even to zero at the input scale. Evolution will
change this and the valence GPD will start to grow also at $X=\zeta$ since
higher and higher Fock states will be present in the valence GPD at
higher $Q^2$ as previously discussed in Sec.\ \ref{origin}. This
prediction is supported by several model calculations. First,
calculations both in the chiral-quark-soliton model \cite{wgp}, in the
constituent quark model \cite{scopetta} and within a light cone
wavefunction approach \cite{dfjk} show that the valence GPD becomes
either small or vanishes at $X=\zeta$.  In the chiral-quark-soliton model,
for example, the contribution to the flavor singlet of the discrete
Dirac spectrum is identified with the bound state quark structure both
valence and sea, whereas the continuum part is identified with the
pion field itself \cite{wgp} or what I termed nonperturbative QCD
vacuum fluctuations.  The continuum part rapidly changes behavior from
an increasing to a decreasing function which is essentially $0$ at the
crossover $X=\zeta$.  This is easily explainable if one remembers that the
asymmetric $q\bar q$ fluctuations from the nonperturbative vacuum at a
low scale correspond to endpoint contributions in the pion or meson
wavefunction which are suppressed. Note that the continuum
contribution is C-even and thus the individual flavor contributions
enter the DVCS amplitude with the square of their respective charges.
The bound state quark distribution has both a C-even and C-odd part,
where the flavor decomposed C-even part contributes to the DVCS
amplitude.  In order to replicate the value of this distribution at
$X=\zeta$ as well as its functional behavior (see Fig.2 of \cite{wgp}),
the value of the C-even and C-odd parts at $X=\zeta$ should be both
positive but smaller than the value of the total bound state quark
distribution.  This means that both the C-even and C-odd or valence
distribution in the DGLAP region have essentially the same functional
behavior as the total distribution.  This means that the sum of the
C-even distribution from the continuum and discrete part as well as
the valence distribution yield a falling distribution towards $X=\zeta$ at
a nonperturbative scale, as I advocate.  In other words, the required
configurations for DVCS are rare at a nonperturbative scale.

Secondly, since $X=\zeta$ corresponds to large light-like separations as
in the inclusive case for $x_{bj}\to0$, one might expect that the
nonperturbative valence quark GPD actually vanishes at $X=\zeta$ as the
forward valence quark PDF vanishes for $x_{bj}\to0$.  Experimentally,
this could be verified in principle through a flavor separation in $\nu
p$ DVCS at the COMPASS experiment at very low $Q^2$ with the two huge
caveats of unknown higher twist and a large BH contribution at very
low $Q^2$ and large $x_{bj}$.

The slope of the $t$ dependence for small values of $t$ in DVCS at low
$Q^2\sim$ a few $\mbox{GeV}^2$ should be larger than the one for light
meson production for the same kinematics, whereas at large $Q^2$, the
two slopes should be the same as stated in factorization theorems
\cite{jcaf,jfs}. The reason for this is quite simple in the region of
$x_{bj}$ and $Q^2$ where the imaginary part of the amplitude
dominates: In meson production, as pointed out above, both asymmetric
quark and gluon configurations couple to the reaction with equal
strength. Whereas in DVCS the coupling strength of the quark and gluon
correlations are very different, $\alpha$ and $\alpha\alpha_s$ respectively. It is
important to note that I do not assume that the asymmetric quark and
gluon configurations have a different spatial distribution in the
transverse plane.

Why is the difference in coupling strength important in this case ?
Because of the difference in coupling strength compared to meson
production the slope in $t$ for DVCS at low $Q^2$ is quark dominated
while in meson production it is a priori a mixture of quarks and
gluons. If quarks and gluons had the same $t$-slope than the difference
in coupling strength would not matter since percentage wise the
amplitude for DVCS and meson production would change the same way in
$t$ and the $t$ slopes would be independent of $Q^2$.  If gluons had a
larger slope in $t$ than quarks, one would expect that for large
$Q^2$, due to the mixing of quarks and gluons under perturbative
evolution which equilibrates the slopes of quarks and gluons, the
slope for DVCS or meson production would increase with $Q^2$ since the
slope for quarks would increase. Both of the above assumptions are not
what the data indicate (see for example \cite{fmsnew,brho}).  Rather
than a constant slope or an increase, one observes a decrease of the
slope with an increase of $Q^2$.  The fact that the smallest slope is
measured in $J/\psi$ photoproduction which is essentially only sensitive
to the gluon GPD, tells us that quarks and gluons have not only
different slopes in $t$ at $Q^2\simeq M_{J/\psi}^2$ which corresponds to small
transverse distances, but that the slope for gluons is smaller than
that for quarks. Going to even lower scales, the difference in slope
can only increase rather than decrease because of the evolution
argument.  Note once more that I do not refer to any particular
difference between spatial distributions of quarks and gluons.

If one were to take $x_{bj}\leq0.01$, $Q^2\simeq2~\mbox{GeV}^2$, integrate out
$t$ and further assume, for simplicity, that the gluon amplitude which
enters both DVCS and meson production in this kinematic region with a
$-$ sign, is between $30-50\%$ of the quark amplitude at low $Q^2$
modulo coupling effects, then it is a simple exercise to show that
the effective $t$ slope for DVCS is larger than for meson production
Furthermore, it is immediately clear that the difference depends on
the relative difference in coupling strength between quarks and gluons
in DVCS and meson production.

Taking the quark slope to be about $8$ and for gluons to be about $4$
seems to be not unreasonable. Furthermore take $\alpha_s\simeq0.3$ and the gluon
about $50\%$ of the quark. The effective slope for DVCS i.e. for the
square of the amplitude assuming a $t$ dependence in the amplitude of
$e^{B_{q,g}t/2}$ for small $t$, is then about $16/1.79$ compared to
$16/2$ for light meson production which will be mainly $\rho$ production
in this kinematic range. Taking the ratio of effective slopes of meson
production to DVCS gives about $0.8$. The difference in the ratio from
$1$ is entirely due to the difference in the coupling strength between
quarks and gluons in DVCS and meson production respectively.

At large $Q^2$, on the other hand, where there will be many, suitable
configurations, originating almost exclusively from gluons, this
difference in coupling strength becomes unimportant due to the very
large number of suitable configurations which leads to an
equilibration of quark and gluon slopes.  The conclusion for low $Q^2$
is supported by the findings in \cite{fmsnew} where a larger slope for
DVCS at relatively low $Q^2\sim 2-4~\mbox{GeV}^2$ was required to obtain
a good agreement between data and theory than in the case of, for
example, $\rho^0$ production \cite{brho} with a ratio of the effective
slopes of about $0.7-0.8$. That the $t$ slopes for quark and gluons
equilibrate at large $Q^2$ and become universal as predicted by
factorization is seen in the effective slope for $\rho$ production
\cite{brho} rapidly approaching the one for $J/\psi$ production with an
increase in $Q^2$, but not going below that value for even larger
$Q^2$.

\section{Conclusions}
\label{conc}

To summarize once more, I have presented a concise, simple and
intuitive picture of what GPDs mean in the sense of carrying new
information about the three dimensional (two transverse and one light
cone dimension) structure of nucleons (more precisely nucleon to
nucleon transitions) compared to inclusive parton distributions or
form factors. To achieve this I have developed a simple picture
through which type of particle configurations encoded in the GPDs,
DVCS and meson production proceed and that these configurations can
only be correctly identified in exclusive reactions. These
configurations originate mainly from symmetric quark configurations
through perturbative evolution. Furthermore, based on this picture, I
conclude that the unpolarized valence quark GPD at a
nonperturbative scale should be either small compared to a an
inclusive valence PDF at the same $x_{bj}$ or vanish near the
crossover point between ERBL and DGLAP region. I have also made
verifiable, qualitative predictions for DVCS and meson production in
$ep$ and $eA$ collisions such as an early onset of saturation,
different geometric scaling curves for different $t$ values,
determining the sizes of the ``grey'' and ``black'' areas of the
target, stronger nuclear shadowing corrections in the transition
region $0.01<x_{bj}<0.1$ and a difference in the slope of the
$t$-dependence at low $Q^2$ between the two processes, using the above
picture. These predictions/conclusions are already partially supported
by both experimental as well as theoretical observations.

I would like to thank Nikolai Kivel, Mark Strikman and Christian
Weiss, for useful discussions as well as Moskov Amarian, Vladimir
Braun, Einan Gardi, Mark Strikman, and Heribert Weigert for
reading the manuscript.  This work was supported by the
Emmi-Noether grant of the DFG FR-1524/1-2.

\end{document}